\documentclass[fleqn,usenatbib]{mnras}

\usepackage{newtxtext,newtxmath}

\usepackage[T1]{fontenc}

\DeclareRobustCommand{\VAN}[3]{#2}
\let\VANthebibliography\thebibliography
\def\thebibliography{\DeclareRobustCommand{\VAN}[3]{##3}\VANthebibliography}

\usepackage{graphicx}	
\usepackage{amsmath}	
\usepackage{caption}
\usepackage{subcaption}

\def\be{\begin{equation}} 
\def\ee{\end{equation}}

\def\gsim{\lower.5ex\hbox{\gtsima}} 
\def\lsim{\lower.5ex\hbox{\ltsima}} \def\gtsima{$\; \buildrel > \over 
\sim \;$} \def\ltsima{$\; \buildrel < \over \sim \;$} \def\prosima{$\; 
\buildrel \propto \over \sim \;$} \def\gsim{\lower.5ex\hbox{\gtsima}} 
\def\lsim{\lower.5ex\hbox{\ltsima}} 
\def\simgt{\lower.5ex\hbox{\gtsima}} 
\def\simlt{\lower.5ex\hbox{\ltsima}} 
\def\simpr{\lower.5ex\hbox{\prosima}} \def\la{\lsim} \def\ga{\gsim} 
  
 \def\gtsima{$\; \buildrel > \over \sim \;$} 
\def\ltsima{$\; \buildrel < \over \sim \;$} 
\def\gsim{\lower.5ex\hbox{\gtsima}} 
\def\lsim{\lower.5ex\hbox{\ltsima}} 
\def\simgt{\lower.5ex\hbox{\gtsima}} 
\def\simlt{\lower.5ex\hbox{\ltsima}} 
\def\simpr{\lower.5ex\hbox{\prosima}} 
\def\la{\lsim} 
\def\ga{\gsim}

\def\E3{{\cal E}_{\rm g}^{III}}

\def\M*{M_*}

\def\Z*{Z_*}
\def\L*{L_*}

\def\der{{\rm d}}

\def\der{{\rm d}}

\def\der{{\rm d}}
\newcommand{\de}{{\rm d}}

\title[Joint constraints on WDM ]{Warm Dark Matter constraints from the joint analysis of CMB, Lyman-$\alpha$, and global 21~cm data}

\author[Chatterjee \& Choudhury]{
Atrideb Chatterjee$^{1, 2}$\thanks{E-mail:atrideb.chatterjee@iucaa.in},
Tirthankar Roy Choudhury$^{1}$\thanks{E-mail:tirth@ncra.tifr.res.in}
\\
$^{1}$National Centre for Radio Astrophysics, TIFR, Post Bag 3, Ganeshkhind, Pune 411007, India\\
$^{2}$Inter-University Centre for Astronomy and Astrophysics, Post Bag 4, Ganeshkhind, Pune 411007, India\\}

\date{Accepted XXX. Received YYY; in original form ZZZ}

\pubyear{2023}

\begin{document}
\label{firstpage}
\pagerange{\pageref{firstpage}--\pageref{lastpage}}
\maketitle

\begin{abstract}
With the help of our previously built MCMC-based parameter estimation package \texttt{CosmoReionMC}, we investigate in detail the potential of 21~cm global signal, when combined with CMB and observations related to the QSO absorption spectra, to constraint the mass of Warm Dark Matter (WDM) particle. For the first time, we simultaneously vary all the free parameters (mass of WDM particle, cosmological parameters, and astrophysical parameters) in a joint analysis with CMB, observations related to the QSO absorption spectra and 21~cm global signal, to address the long-overlooked issue of the possible degeneracies between the Dark Matter particle mass $m_X$ and cosmological/astrophysical parameters. From the existing CMB and QSO absorption spectra data, we can rule out $m_X < 2.8$~keV at 95\% confidence level. Including a mock 21~cm global signal in the redshift range $z = 25 - 5$ expected to be observed with upcoming instruments designed for global signal, the forecasted constraint is found to be much tighter $m_X > 7.7$~keV, assuming that the true dark matter model is the usual cold dark matter. In case the mock 21~cm signal is constructed for dark matter particles having $m_X = 7$~keV, our forecasts indicate that $\left(m_X / \text{keV}\right)^{-1}$ is in the range $[0.1, 0.2]$ ($95\%$ confidence level). This implies that the future 21~cm data should allow detection of the WDM particle mass if $m_X \sim 7$~keV.

\end{abstract}




\section{Introduction}

The concordance $\Lambda$CDM model is extremely successful in explaining the Universe's large-scale structure, e.g., extremely accurate prediction of the cosmic microwave observations \citep{1982ApJ...263L...1P} and large-scale distribution of the galaxies \citep{1984Natur.311..517B}. Interestingly, the same cosmological model fails to match with some of the galactic and sub-galactic scale observations such as - $(i)$ Dearth of low mass galaxies in low mass halos \citep{2001ApJ...557..495P}, $(ii)$ $\Lambda$CDM model's prediction of cuspy core in the DM halos opposes the observationally preferred constant density cores \citep{moore1999b, subramanian2000}, $(iii)$ Too big to fail problem for the field galaxies \citep{2016MNRAS.460.3610O}. The root cause behind all of these problems is the abundance of the small-scale structure due to the very cold nature (with mass $\sim$ 100 GeV) of the constituent DM particle in the $\Lambda$CDM model.

Recently, a number of hydrodynamical simulations \citep{2004ApJ...609..482K, 2010Natur.463..203G, 2014AAS...22331006T, 2016MNRAS.457.1931S, 2019MNRAS.489.4574G, 2021MNRAS.507.4211E, 2021ApJ...906...96A, 2021JCAP...12..046G} have been trying to solve this issue by considering the baryonic feedback in the form of AGN, or stellar feedback to inhibit the overproduction of the small scale structure. Nevertheless, incorporating the baryonic feedback self consistently with the DM-only simulation is extremely non-trivial, and so far, the success is limited \citep{2008MNRAS.390..920O, boylan2011, 2012MNRAS.422.1231G, 2013MNRAS.429.3068T}. An alternate solution to these crises that has been proposed is to assume that the DM is ``non-cold'' \citep{boehm2001, wang2014, 2000PhRvL..85.1158H}. The generic feature of such DM candidates is that the small-scale fluctuations in the matter distribution are suppressed relative to the standard CDM. There are several examples, for instance, Warm Dark Matter like sterile neutrinos \citep{PhysRevLett.72.17, Laine_2008, 2016MNRAS.461...60L}, ultra-light scalars or axions also known as Fuzzy Dark Matter \citep{2000PhRvL..85.1158H, 2014MNRAS.437.2652M, du2017, 2022PhRvD.105h3011G}, self-interacting Dark Matter \citep{PhysRevLett.84.3760, vogelsberger2014_wdm} and others \citep{boehm2001, wang2014, dvorkin2014}.

Among the above, one of the most extensively studied candidates is the Warm Dark Matter (WDM) with particle masses $m_x \sim \mathcal{O}$~keV \citep[see, e.g.][]{blumenthal1984, bode2001, 2012NewA...17..653D, 2012MNRAS.420.2318L}. These particles are essentially thermal relics; hence, the small-scale suppression is entirely determined by particle mass \citep{2012MNRAS.420.2318L}. Since these models have been widely studied in the literature, there exist straightforward methods to compute the abundance of dark matter haloes in addition to the modifications in the DM power spectrum \citep{bode2001, viel2005, 2014MNRAS.439..300L, schneider_wdm2014, 2020ApJ...897..147L}, both of which are crucial for our work. 
 
Given the lower value of the WDM particle mass, these models erase the small-scale substructure, delay the structure formation and therefore solve the small-scale problems arising in the $\Lambda$CDM model.

As the WDM models delay the structure formation, it consequently delays the formation of the first stars. Therefore any observation related to the formation of the first stars could be used to constrain the mass of the WDM particles. As reionization is believed to start from the first generation of stars, a number of studies \citep{2001ApJ...558..482B, 2003ApJ...598...73Y, 2003ApJ...593..616S, 2012ApJ...747..127Y, 2013MNRAS.435L..53P, dayal2015, dayal2017a, 2017PhRvD..96j3539L, 2021MNRAS.507.3046R, 2023PhRvD.108d3030S} used reionization related observations to put lower limits on the mass of the WDM in the range of $1.3 - 5$ keV.  Another observation related to the first generation of stars is the global 21~cm signal coming from cosmic dawn. A growing body of studies \citep{2019PhRvD.100l3005B, 2020JCAP...04..004L, 2020MNRAS.497.3393R, safarzadeh2018, chatterjee2019, 2022ApJ...929..151H} exploited this signal to put lower limits on the WDM particle mass in the range of $3 - 6.6$ keV. Furthermore, \cite{2013PhRvD..88d3502V, irsic2017} and more recently \cite{PhysRevD.98.083540} used Lyman Alpha (Ly$\alpha$) forest power spectrum measurement using MIKE/HIRES spectrograph coming from high-resolution quasar spectra at redshifts z $\sim 2 - 5$ to constrain the WDM mass in the range $2.2- 3.6$ keV \footnote{Note that the exact limit on $m_{X}$ will depend on the priors taken regarding the IGM temperature and the choice of the IGM temperature evolution model.}. Very recently, the high redshift observations coming from the James Webb Space Telescope (JWST) have been used to rule out WDM models with $m_{X} < 1.5-2.0$ keV \citep{2022arXiv221103620M, 2023arXiv230314239D}. Other than these high redshift observations, \cite{kennedy2014} used the count of dwarf galaxies to rule out WDM mass $m_{X} < 2.3$ keV, and finally, the most stringent constraint on the WDM mass comes from \cite{2021ApJ...917....7N} ruling out $m_{X} < 9.7$ keV from a combined analysis of strong gravitational lenses and the Milky Way satellite galaxy population. 

However, the aforementioned works have two limitations: $(i)$ All of these works use either the reionization-related observations, the global 21~cm signal, or other high-redshift observations, but none of them combines all the data to put constraints on the WDM particles. $(ii)$  Some of the above-mentioned works, where hydrodynamical or semi-numerical simulation is used, did not employ MCMC-based methods to quantify degeneracies (if any) between the mass of WDM particle and other cosmological and/or astrophysical parameters. For example, if we change any cosmological parameter that can delay the timing of structure formation of the Universe, that can, in principle, imitate the effect of lowering the mass of a WDM particle (remember that the lighter the mass of a WDM particle, the harder it is to start the structure formation).

To overcome both these issues, one has to first combine CMB, reionization-related observations and a hypothetical dataset of 21~cm signal and then vary all the free parameters (mass of WDM particles along with all the other cosmological and astrophysical parameters) simultaneously to put constraints on the mass of WDM particles and quantify, if any, degeneracy between different free parameters. In \cite{2021MNRAS.507.2405C} (referred to as CCM21 hereafter), we have introduced an advanced MCMC-based parameter estimation package called \texttt{CosmoReionMC} which has all the above-mentioned features and therefore provides an ideal opportunity to carry out this investigation.

The rest of the paper is organised as follows. We describe in Section-\ref{Theo_model} the effect of incorporating the WDM in our galaxy formation model, theoretical modelling for reionization and 21~cm signal. Section-\ref{sec:result} describes the findings of this work, and finally, Section-\ref{conclude} summarizes the work.

\section{Theoretical Modelling}
\label{Theo_model}
\subsection{Warm dark Matter}\label{wdm_modification}
It is well known that the effect of introducing the mass of WDM in our reionization and global 21~cm signal modelling will be manifested in the DM Power spectrum, the halo mass function and consequently on any quantity that depends on either or both of them.

Following \cite{bode2001}, the DM power spectrum of the WDM can be expressed as
\begin{equation}
    P_{\rm WDM}(k)=T_{\rm WDM}^2(k)P_{\rm CDM}
\end{equation}
where $P_{\rm CDM}$ is the usual CDM power spectrum and $T_{\rm WDM}$ is the transfer function given by \citep{viel2005}
\begin{equation}
    T_{\rm WDM}=\left[ 1+ (\alpha k)^{2\mu} \right]^{-5/\mu}
\end{equation}
where $\mu=1.2$ and $\alpha$ is given by \citep{viel2005}
\begin{equation}\label{eqn:alpha_eq}
    \alpha=0.049\left( \frac{m_{\rm X}}{\rm keV}\right)^{-1.11} \left(\frac{\Omega_{\rm WDM}}{0.25} \right)^{0.11} \left( \frac{h}{0.7} \right)^{1.22} \,\,\,\, \rm Mpc/h
\end{equation}
 Following \cite{2020ApJ...897..147L}, we write the halo mass function of the WDM  as
\begin{equation}
    \frac{\partial N_{\rm WDM}}{\partial  M}=\left[ 1+ \left(\beta\frac{ M_{\rm hm}}{M}\right)^{\gamma} \right]^{\delta} \frac{\partial N_{\rm CDM}}{\partial M}
\end{equation}

 with $\beta= 2.3$  $\gamma = 0.8$, $\delta = -1.0$ $ N_{\rm CDM}$ and $N_{\rm WDM}$ are the number of CDM and WDM haloes respectively. The half-mode mass $M_{\rm hm}$ is given by
\begin{equation}
    M_{\rm hm}=\frac{4}{3} \pi \Bar{\rho}_{m} \left(\frac{\lambda_{\rm hm}}{2} \right)^{3}
\end{equation}
$\Bar{\rho}_{m}$ is the background matter density, and the half-mode scale is given by
\begin{equation}
    \lambda_{\rm hm}=2 \pi \alpha \left(2^{\mu/5}-1 \right)^{-1/2\mu} \,\, \rm Mpc/h
\end{equation}
 Following \citep{sheth-tormen1999}, the halo mass function for the CDM is given by
\begin{equation}
    \frac{\partial N_{\rm CDM}}{\partial M}=-\frac{1}{2} \frac{\Bar{\rho}_{m}}{M^2} \frac{d \log \sigma^{2}}{d\log M} f(\nu) 
\end{equation} 
  where 
\begin{equation}
    f(\nu)=A\sqrt{\frac{2q\nu}{\pi}} \left[ 1+ \left(q \nu \right)^{-p} \right]\exp \left[ -\frac{q\nu}{2} \right]
\end{equation}
with $A = 0.3222$, $q = 0.707$, $p = 0.3$ and $\nu$ is defined as
\begin{equation}
    \nu=\frac{1.686}{\sigma^{2}(M)D(z)}
\end{equation}
where $D(z)$ is the well-known growth function.

\subsection{Modelling Reionization}\label{reion_modification}

For this work, we use the reionization model (CF model hereafter) developed in \cite{2005MNRAS.361..577C, 2006MNRAS.371L..55C, 2012MNRAS.419.1480M}. As the detailed discussion of this model is beyond the scope of this paper, we briefly summarize here the main characteristics of this model. 
\begin{itemize}
    \item In this model, the overdensity of the intergalactic medium (IGM) is described using a lognormal distribution in the low-density regions and as a power law distribution in the high-density region following the treatment presented in \cite{2003ApJ...597...66M}. We write the probability density function (PDF) of the overdensity $\Delta$ as
    \begin{align}
        P(\Delta) &= \frac{A}{\sigma_b \Delta \sqrt{2 \pi}} \exp\left[-\frac{(\ln \Delta - \mu)^2}{2 \sigma_b^2}\right] & \mathrm{if} \Delta < \Delta_V,
        \nonumber \\
        & = \frac{A}{\sigma_b \Delta_V \sqrt{2 \pi}} \exp\left[-\frac{(\ln \Delta_V - \mu)^2}{2 \sigma_b^2}\right] \left(\frac{\Delta}{\Delta_V} \right)^{\beta} & \mathrm{if} \Delta > \Delta_V,
    \end{align}
    where the parameters $A$, $\mu$ and $\Delta_V$ are determined by demanding continuity of the derivative of $P(\Delta)$ at the transition overdensity $\Delta_V$, and by normalizing the volume and mass to unity. We choose $\beta = -2.5$, appropriate for high redshifts. The quantity $\sigma_b$ is the rms linear mass fluctuations in baryons and is related to the WDM power spectrum as
    \begin{equation}
        \sigma_b^2 = \int_0^{\infty} \mathrm{d}k~k^2~ \frac{P_\mathrm{WDM}(k)}{\left(1 + x_J^2 k^2\right)^2}
    \end{equation}
    where $x_J$ is the Jeans length, which depends on the IGM temperature. Note that the density PDF is sensitive to the value of $m_X$ through $\sigma_b$.   
    The most important feature of this model is its ability to calculate the ionization and thermal state of the IGM in the neutral and ionized regions for different species (i.e., hydrogen and helium) separately, simultaneously and self-consistently. Moreover, once all the low-density regions of the IGM are ionized, this model assumes the Universe to be completely ionized. 
    \item The original CF reionization model \citep{2005MNRAS.361..577C, 2006MNRAS.371L..55C, 2012MNRAS.419.1480M} assumes the source of reionization to be quasars, PopII and PopIII stars. Therefore the total photon production rate at a redshift z is given by
\begin{equation}\label{eqn:n_photon}
    \dot{n}_{\rm ph} (z) = \dot{n}_{\rm ph, \rm stellar}(z)+ \dot{n}_{\rm ph, \rm QSO}(z)
\end{equation}
While the quasar contribution can be calculated easily by computing their ionizing emissivities from the observed quasar luminosity function (LF) at $z<7.5$ \citep{2019MNRAS.488.1035K}, the calculation for stellar contribution becomes non-trivial if we consider both PopII and PopIII. However, CCM21 shows that the contribution of PopIII stars is negligible as long as we use CMB and quasar absorption-related observations to constrain different parameters. Also, as discussed later, while simulating the global 21~cm signal, we take into account the contribution only from PopII stars. Therefore, in this work, we take stellar contributions only from PopII stars and completely ignore the contributions from PopIII stars. The number of ionizing photon from stellar sources are hence computed using
\begin{equation}
    \dot{n}_{\rm ph, \rm stellar}(z)= \rho_b \mathbf{\epsilon} \frac{\der f_{\mathrm{coll}}}{\der t} \int^{\infty}_{\nu_H}  \left(\frac{\der N_{\nu}}{\der M} \right) \der \nu
\label{eqn:photon_production}
\end{equation}
where $\nu_{H}$ is the threshold frequency for hydrogen photoionization, $\rho_b$ is the mean comoving density of baryons in the IGM, and  $\epsilon = f_{*} \times f_{\mathrm{esc}}$, where $f_*$ and $f_{\mathrm{esc}}$ respectively denotes the star formation efficiency and the escape fraction of the ionizing photons. The quantity $\der N_{\nu}/ \der M$ denoting the number of photons emitted per frequency range per unit mass of the star, depends on the stellar spectra and IMF of the stars \citep{2005MNRAS.361..577C}. Using a standard Salpeter IMF in the mass range $1-100 M_{\odot}$ with a metallicity of $0.05 M_{\odot}$, $\der N_{\nu}/ \der M$ has been computed from the stellar synthesis models of \citet{2003MNRAS.344.1000B}. We consider $\epsilon$ as a free parameter in our model and later constrain it using MCMC (discussed in Sections 3 and 4).

\item Two of the observables that the CF model can predict and will be later used in our MCMC analysis are $(i)$ the redshift distribution of Lyman-limit system $(dN_{\rm LL}/dz)$ and $(ii)$ the hydrogen photoionization rate $(\Gamma_{\rm PI})$. To calculate both the observable, the CF model first calculates the mean free path of the photons  using 
\begin{equation}
\label{eq:equation_3}
    \lambda_{\mathrm{mfp}}=\frac{\lambda_{0}}{[1-F_{v}(\Delta_{i})]^{2/3}}
\end{equation}
where $\lambda_{0}$ is a free parameter of the reionization model, and 
\begin{equation}
    F_{V}(\Delta_i) = \int_0^{\Delta_i} \mathrm{d}\Delta ~ P(\Delta)
\end{equation} 
is the volume fraction of the ionized region as a function of the overdensity $\Delta_{i}$. It is clear that the mean free path is sensitive to $m_X$ through $P(\Delta)$. The dependence of $\lambda_{\rm mfp}$ on the mass of the WDM particle is described in detail in Appendix-\ref{appendix: mfp}. Once we compute $\lambda_{\rm mfp}$, it is straightforward to calculate  $dN_{\rm LL}/dz$ \citep[CCM21]{2005MNRAS.361..577C}. Similarly, the $\Gamma_{\rm PI}$ can be calculated using 
 \begin{equation}
      \Gamma_{\mathrm{PI}}(z)=(1+z)^{3} \int^{\infty}_{\nu_H} \de \nu~ \lambda_{\mathrm{mfp}}(\nu,z)~\dot{n}_{\mathrm{ph}}(z)~\sigma_{H}(\nu),
 \end{equation}
 where $\sigma_{H}(\nu)$ is the hydrogen photoionization cross-section and $\dot{n}_{\mathrm{ph}}(z)$ is the photon production rate as described in eqn-\ref{eqn:n_photon}.
\end{itemize}

\begin{figure}
    \centering
    \includegraphics[width=\columnwidth]{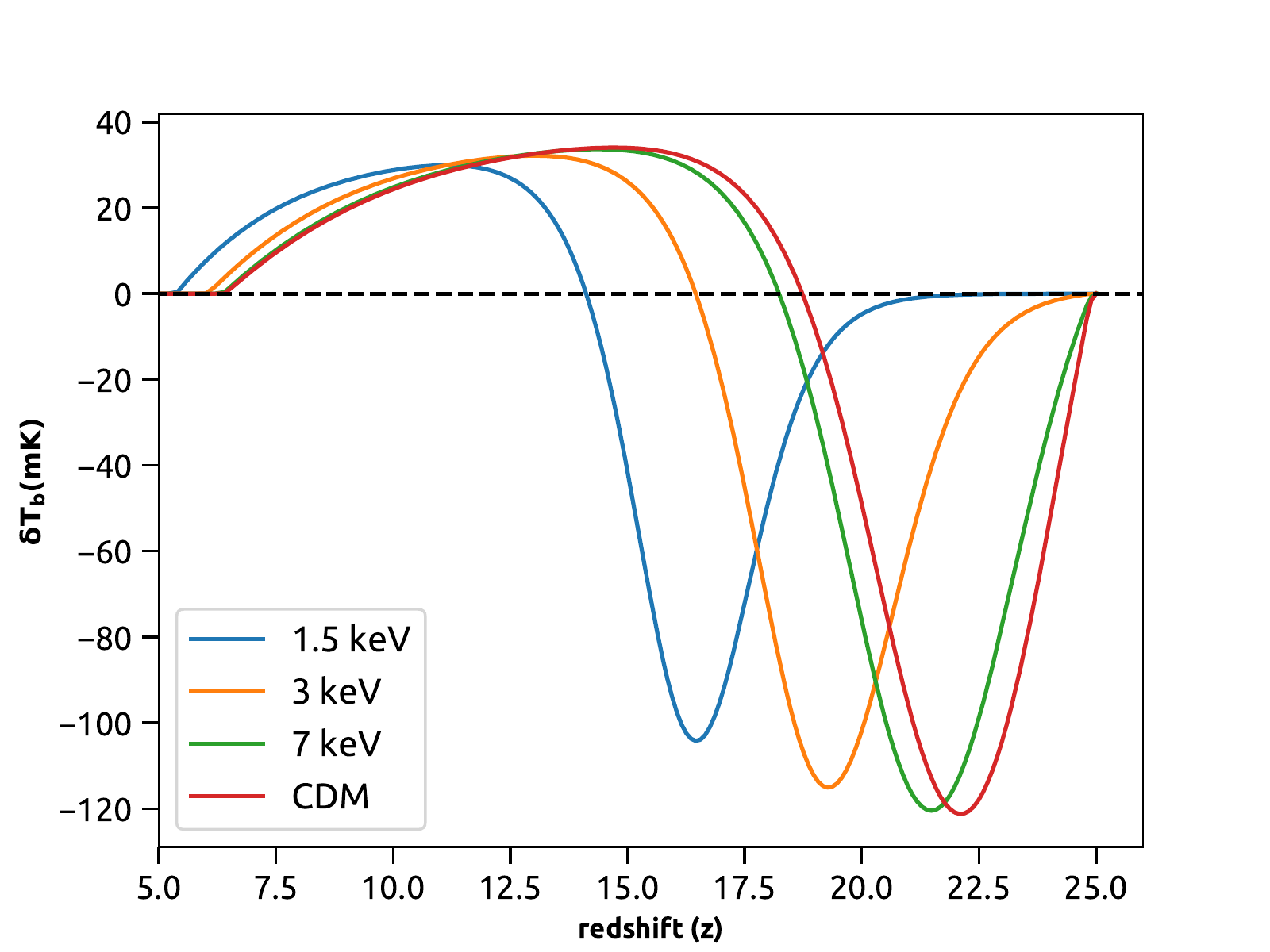}
    \caption{The global 21~cm differential brightness temperature for the CDM, 7 keV, 3 keV and 1.5 keV WDM models. As is evident, if we lower the mass of the DM particles, the absorption trough in the signal shifts to lower redshifts.}
    \label{fig:21_WDM}
\end{figure}
Although the introduction of the WDM models does not change the mathematical framework used for the CF reionization model, it affects any quantity of the model that depends on the halo mass function (hmf). For example, the collapse fraction of the DM halo (appeared in eqn-\ref{eqn:photon_production}), which depends on the hmf, will change and the modified form will be
    \begin{equation}
        f_{\mathrm{coll}} = \frac{1}{\Bar{\rho}_{m}}\int^{\infty}_{M_{\mathrm{min}}(z)} \de M  M \frac{\partial N_{\rm WDM}(M,z)}{\partial M},
    \end{equation}
where $\Bar{\rho}_{m}$ is the mean comoving density of dark matter, $M_{\mathrm{min}}(z)$ is the minimum mass for star-forming halos which is determined by different cooling processes (such as atomic cooling, molecular cooling) and feedback processes (radiative feedback, mechanical feedback, chemical feedback, Lyman Warner feedback etc.). In the reionization model considered here, we consider only the atomic cooling and on top of that, radiative feedback is incorporated using a Jeans mass prescription described in detail in \cite{2005MNRAS.361..577C}.
    
\subsection{Global 21~cm modelling}\label{global_signal} 

The sky averaged  21~cm global differential brightness temperature can be written as \citep{furlanetto2006c, chatterjee2019}
\begin{equation}
    \delta T_b(\nu) \approx 10.1~\mathrm{mK}~x_{\mathrm{HI}}(z) \left[1 - \frac{T_{\gamma}(z)}{T_S(z)}\right]~(1+z)^{1/2},
\end{equation}
where $T_{\gamma}$ is the background radiation temperature, $T_{S}$ is the neutral hydrogen spin temperature and $x_{\mathrm{HI}}$ is the neutral hydrogen fraction in the IGM. Under the assumption that the optical depth of the Ly${\alpha}$ is very high in the epoch of cosmic dawn and the redshift range we are interested in (discussed later in Section-\ref{constrain_with21cm}), the spin temperature $T_{S}$ is computed using
\begin{equation}
    T_S^{-1}=\frac{T_{\gamma}^{-1} + x_{\alpha} T_K^{-1}}{1+x_{\alpha}}.
\end{equation}
where $T_{K}$ is the kinetic temperature of the IGM, $x_{\alpha}$ is the Ly$\alpha$ is the
coupling coefficient \footnote{In the redshift range of our interest, i.e. $z = 25 - 5$, the collisional coefficient is insignificant as the number density of the free electrons and protons becomes negligible due to the expansion of the Universe \citep{2012RPPh...75h6901P}}.

Although the kinetic temperature of the IGM computed in the CF reionization model is decided mainly by two processes, namely, the adiabatic cooling and the photoheating from UV photons, the moment we consider a 21~cm signal coming from cosmic dawn, we have to include the X-ray heating term in the temperature evolution equation. However, as we go towards lower redshift during EoR, the X-ray heating can be ignored, and the UV heating becomes dominant once the reionization starts; therefore, we turn off the X-ray heating in the reionization epoch.

The X-ray heating can be computed using \citep{2012MNRAS.419.2095M}
\begin{equation}
\frac{\epsilon_{X}}{\mathrm{J}~\mathrm{s}^{-1} \mathrm{Mpc}^{-3}} = 3.4 \times 10^{33} \frac{\rho_{b} f_{Xh, *} \frac{\de f_{\mathrm{coll}}}{\de t} }{M_{\odot} \mathrm{Mpc}^{-3} \mathrm{yr}^{-1}},
\end{equation}
where $f_{Xh, *}=f_{X} \times f_{h} \times f_{*}$. $f_{X}$ is an unknown normalization parameter of our model. It takes into account any discrepancy between the properties of the locally observed galaxy and yet-to-observe high-redshift galaxy. $f_{h}$ is the parameter corresponding to the fraction of the total X-ray photons that heat the IGM.

To calculate $x_{\alpha}$, we first determine the background Ly$\alpha$ flux using
\begin{equation}
J_{\alpha}(z) =\frac{c}{4 \pi}(1+z)^3 \int^{z_{\mathrm{max}}}_{z} \de z'~f_{\alpha, *}~\dot{n}_{\nu'}^{\mathrm{II}}(z')  \left|\frac{\de t'}{\de z'}\right|,
\end{equation}
where $f_{\alpha, *}=  f_{\alpha} \times f_{*}$. $f_{\alpha}$ is an unknown efficiency parameter such that any uncertainties in the properties of the high redshift galaxies can be absorbed in this. Further, the effect of any radiative cascading, generating any additional Ly$\alpha$ photons, will also be absorbed in this factor.\footnote{As the mean free path of X-ray photons is large, it will affect the Ly$\alpha$ flux only far from sources as discussed in \cite{2012RPPh...75h6901P}. Therefore we ignore the effect of X-ray heating while calculating the Ly$\alpha$ background.}
To determine the upper limit $z_{\mathrm{max}}$ of the integral, we assume that all the continuum ionizing photons would be absorbed in the IGM and will not play any part in determining the Ly$\alpha$ radiation.
$z_{\mathrm{max}}$ is calculated using \citep{2020MNRAS.496.1445C} 
\begin{equation}
    1+z_{\mathrm{max}}=\frac{\nu_{H}}{\nu_{\alpha}} (1+z),
\end{equation}
where $\nu_{\alpha}$ is the Ly$\alpha$ frequency.\footnote{Note that we have ignored here the Ly$\alpha$ heating of the IGM \citep[see, e.g.,][]{2020MNRAS.492..634G}.}
The quantity $\dot{n}_{\rm \nu'}(z')$ is given by
\begin{equation}
    \dot{n}_{\rm \nu'}(z')= \rho_b \frac{\der f_{\mathrm{coll}}}{\der t}  \left(\frac{\der N_{\nu'}}{\der M} \right) 
\label{eqn:nu_production}
\end{equation}
with $\frac{\der N_{\nu'}}{\der M}$ denoting number of photon per unit stellar mass at frequency $\nu'$.
Once we determine the background Lyman-$\alpha$ flux, the coupling coefficient can be computed using 
\begin{equation}
    x_{\alpha} = 1.81 \times 10^{11} (1+z)^{-1} S_{\alpha} \frac{J_{\alpha}(z)}{\rm cm^{-2}s^{-1} Hz^{-1}sr^{-1}}
\end{equation}
where $\rm S_{\alpha}$ accounts for the detailed atomic physics involved in the scattering process, and we take $S_{\alpha} = 1$  \citep{furlanetto2006c}.

\subsubsection{Effect of WDM on 21~cm signal}\label{global_signal_WDM} As discussed earlier, any quantity that depends on the mass of the WDM particle will change due to the change in the mass of the WDM particle and, therefore, the global 21~cm signal as a whole will depend on the value of the WDM mass. In figure-\ref{fig:21_WDM}, we have shown the effect of changing WDM mass on the global 21~cm signal. It is evident from figure-\ref{fig:21_WDM} that as we keep decreasing the mass of DM particles, the absorption trough of the global signal continues to shift towards lower redshifts. This is expected as the smaller value of DM particles will lead to a delay in structure formation and therefore cause the absorption trough of the global signal to occur in lower redshifts.

\subsection{The CMB anisotropies}
While describing the calculation of the CMB anisotropies in CCM21, we mentioned that we modify the publicly available python-wrapped CAMB \cite{Lewis:2013hha}\footnote{\href{https://camb.readthedocs.io/en/latest/}{https://camb.readthedocs.io/en/latest/}} to incorporate the reionization history implied by the CF reionization model rather than using the default redshift symmetric tanh model in the CAMB code. For this work also, we use the same modified CAMB code to generate the CMB anisotropy data.

\subsection{The \texttt{CosmoReionMC} Package}
As the mass of the DM particle is considered as a free parameter for this work, we have modified our previously developed MCMC-based parameter estimation based package \texttt{CosmoReionMC} according to the methods described in Sections - \ref{wdm_modification}, \ref{reion_modification}. Moreover, the version of the \texttt{CosmoReionMC} used here is more flexible compared to the original version as the inverse of the mass of WDM particles i.e., $m^{-1}_{X}$ is treated as a free parameter. 

Next, we will describe the results obtained from this study.

\section{Result}\label{sec:result}
Here we present the results of our analysis on the parameter constraints obtained using \texttt{CosmoReionMC}.

\subsection{Constraining mass of the WDM particles using CMB and Quasar absorption data}

Using \texttt{CosmoReionMC}, we first obtain constraints on the mass of WDM particles using CMB and reionization-related observations while simultaneously varying the cosmological and astrophysical parameters. The free parameters for this analysis (referred to as \textbf{CMB+Quasar} hereafter) are
\begin{equation}
    \Theta =\{H_{0}, \Omega_{b}h^{2}, \Omega_ch^{2}, A_{s}, n_s, \epsilon, \lambda_{0}, m^{-1}_{X} \},
\end{equation} 
where the first five parameters are the usual cosmological parameters, $\epsilon$, $\lambda_{0}$ are the free parameters of our reionization model, and $m^{-1}_{X}$  is the inverse of the mass of the DM particles in the unit of $\rm keV^{-1}$.

In this analysis, we include the reionization-related observations from quasar absorption spectra and the Planck 2018 observations.  Data sets related to reionization used in this analysis are $(i)$ photoionization rate $\Gamma_{\mathrm{PI}}$ data obtained from the combined analysis of quasar absorption spectra and hydrodynamical simulations  \citep{2013MNRAS.436.1023B, 2018MNRAS.473..560D,2011MNRAS.412.2543C, 2021MNRAS.508.1853B}. $(ii)$ The redshift distribution of Lyman-limit system $\de N_{\mathrm{LL}} / \de z$ \citep{2011ApJ...736...42R, 2013ApJ...765..137O, 2013ApJ...775...78F,  2010ApJ...718..392P, 2019MNRAS.482.1456C, 2010ApJ...721.1448S}, $(iii)$  Measurement of the upper limit on the neutral hydrogen fractions coming from the dark fractions in quasar spectra \citep{2023ApJ...942...59J} have been used as priors while calculating the likelihood. On top of that, with the recent studies of the large-scale fluctuations of the effective Ly$\alpha$ optical depth from high redshift quasar spectra \citep{2015MNRAS.447.3402B,2018MNRAS.479.1055B,2017ApJ...840...24E,2018ApJ...864...53E, 2020arXiv200308958C}, we put a prior that reionization has to be completed ($Q_{\mathrm{HII}} = 1$) at $z \geq 5.3$.

The total Likelihood function for this analysis is given by,
\begin{equation}
    \mathcal{L}=\mathcal{L}_{\mathrm{Pl}} + \mathcal{L}_{\mathrm{Re}},
\end{equation}
where
\begin{equation}
\mathcal{L}_{Re}=\frac{1}{2}\sum^{N_{\mathrm{obs}}}_{\alpha=1}\left[\frac{\zeta^{\mathrm{obs}}_{\alpha}-\zeta^{\mathrm{th}}_{\alpha}}{\sigma_{\alpha}}\right]^2.
\end{equation}
Here $\zeta^{\mathrm{obs}}_{\alpha}$ represents the set of $N_{\mathrm{obs}}$ observational data related to photoionization rates and the distribution of the Lyman-Limit system whereas $\zeta^{\mathrm{th}}_{\alpha}$ represents the values from the theoretical model. The $\sigma_{\alpha}$ denotes the observational error bars. $\mathcal{L}_{\mathrm{Pl}}$ is the log-likelihood function corresponding to the Planck 2020 observations \citep{2020A&A...641A...6P}.

We assume a broad flat prior for all the eight free parameters. For $m^{-1}_{X}$ (in unit of $\rm keV^{-1}$), we take the flat prior range to be $[0.0, 1.0 ]$ which allows us to explore $m_X$ in the range $[ \infty, 1.0 ]$ keV.  In order to explore the parameter space with MCMC chains, we use 32 walkers taking $10^{6}$ steps. Before producing any result, the convergence of the MCMC chains is ensured using a detailed auto-correlation analysis as described in CCM21 and \cite{2013PASP..125..306F}.

\begin{figure}
    \centering
    \includegraphics[width=\columnwidth]{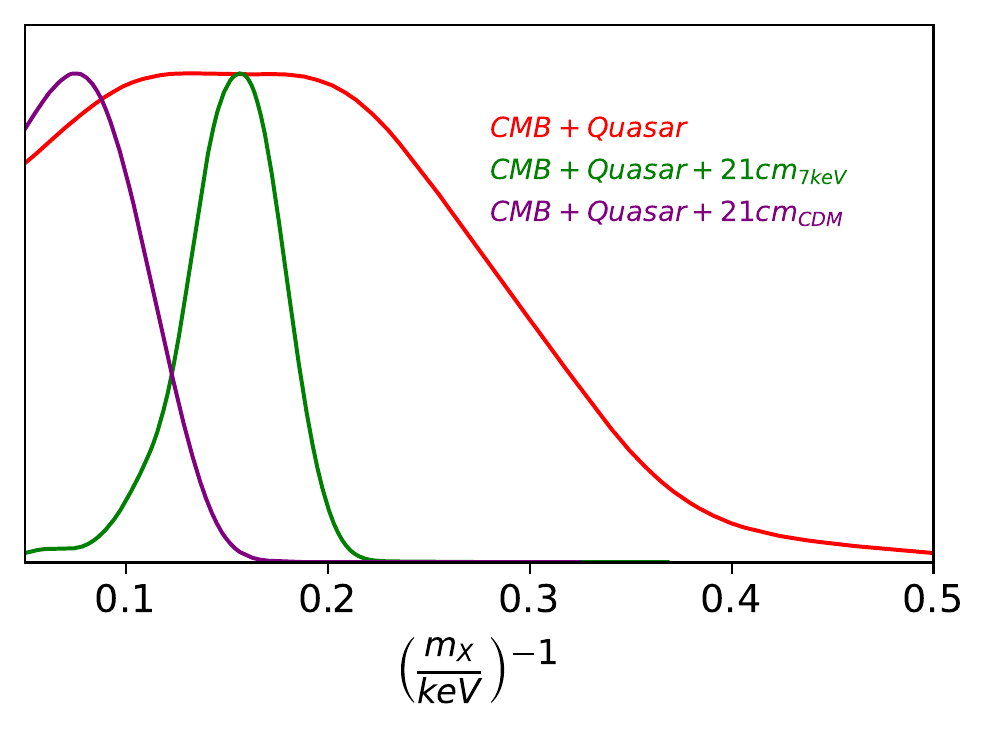}
    \caption{1D marginalized posterior distribution of the constraints on $m^{-1}_{X}$ from different scenarios. The red curve represents the case when CMB and Quasar observations are used. The magenta curve depicts the scenario when the added 21 cm signal is simulated from the CDM model, whereas the green curve represents the case with the hypothetical 21 cm signal computed from the 7 keV WDM model.}
    \label{fig:compare_mx}
\end{figure}

The 1D marginalized posterior distribution of the $\left(\frac{m_{X}}{keV}\right)^{-1}$ is shown (in red) in figure-\ref{fig:compare_mx}. This figure shows that this analysis rules out WDM particles with $m_{X}<2.8$ keV at $95\%$ confidence level. The result is in close agreement with the constraints coming from \cite{baur2016} (they constrain $m_{X} <2.96 $ keV  at $95\%$ confidence level.) However, it is slightly weaker than that of \cite{2013PhRvD..88d3502V} (they rejected WDM particles with $m_{X}<3.3$ eV) and is comparable with the constraints coming from \cite{PhysRevD.98.083540}. Based on the combined observations of medium-resolution spectra of the XQ-100 sample observed with the X-shooter spectrograph ($z \sim 4$) and high-resolution spectra of the $z \sim 5$ QSOs obtained with the HIRES/MIKE spectrographs, \cite{PhysRevD.98.083540} rejected the WDM particles with $m_{X}< 2.2 - 4.1$~keV (at $95\%$ confidence level).  The exact constraint on WDM mass depends on their assumption regarding the temperature evolution of the IGM. As discussed in \cite{PhysRevD.98.083540}, one has to keep in mind that these constraints on the WDM mass also depend on the choices of the priors on the IGM thermal history and that the different priors can significantly alter these mass limits. 

\begin{figure*}
    \centering
    \includegraphics[width=\textwidth]{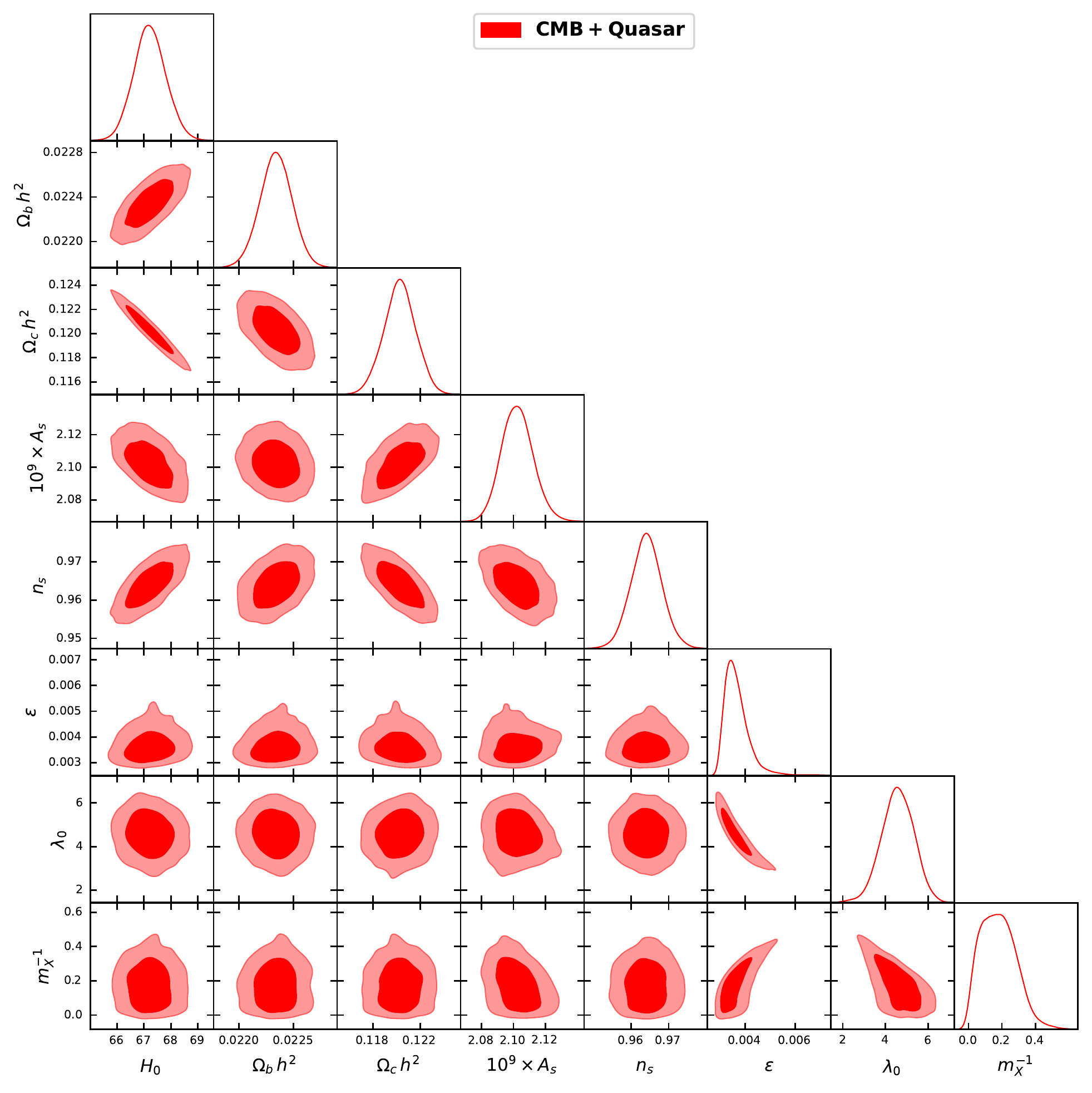}
    \caption{The marginalized posterior distribution of 8 free parameters obtained for the CMB+Quasar case. Two-dimensional plots in the figure show the joint probability distribution (confidence contours at $68\%$ and $95\%$ ) of any two parameters. It is also clear from the two-dimensional plots between $m^{-1}_{X}$ and other free parameters that there exists no correlation between them.}
    \label{fig:WDM_QSO_CMB}
\end{figure*}

For a detailed understanding of the constraints on different parameters and their correlation, the posterior distribution of different parameters is shown in figure- \ref{fig:WDM_QSO_CMB}.  The most important point to note from this figure is that the inverse of WDM mass $m^{-1}_{X}$ has a correlation with $\epsilon$ (and anti-correlation with $\lambda_{0}$). It is clear from $m^{-1}_{X}-\epsilon$ subplot in figure- \ref{fig:WDM_QSO_CMB} that a larger value of $m^{-1}_{X}$ (lower value of $m_{X}$) requires a larger value of $\epsilon$. This is only to be expected because a larger value of $m^{-1}_{X}$ implies a delayed structure formation, and the only way to compensate for this is to have a higher $\epsilon$ to enhance the reionization process and therefore match with the observations. It is also important to note that this correlation is more prominent for a higher value of  $m^{-1}_{X}$ (smaller value of $m_{X}$). This is due to the that a smaller value of $m^{-1}_{X}$ is practically indistinguishable from CDM.  Apart from this, $\epsilon$ and $\lambda_{0}$ shows strong anti-correlation, this is because  $\lambda_{0}$ and $\epsilon$ comes as a product at the time of calculating  $\Gamma_{\mathrm{PI}}$. So,
to keep their product unchanged (necessary to match with the observations), if one parameter increases, the other parameter has to decrease and vice-versa. As $\lambda_{0}$ and $\epsilon$ are strongly correlated and we have already seen that $m^{-1}_{X}$ and $\epsilon$ are correlated, it is only to be expected that $\lambda_{0}$ and $m^{-1}_{X}$ will be anti-correlated.

\subsection{Constraining WDM with CMB, Quasar and a hypothetical global 21~cm signal}\label{constrain_with21cm}

Next, we focus on seeing the potential of the global 21~cm signal to put tighter constraints on the mass of WDM particles when added along with the CMB and Quasar absorption data. To this aim, we first generate a mock 21~cm signal and then add it with the CMB and Quasar data.  While constructing the hypothetical signal, we follow the procedure outlined in CCM21. To briefly summarize, we generate the mock signal in the frequency range 55-235 MHz with frequency channels of width 0.5 MHz. In each frequency channel, we add a Gaussian noise of zero mean and standard deviation $\sigma_{i} = 10$mK to the theoretical signal. Note that this assumed noise is lower compared to what was found in the EDGES or SARAS experiment but is certainly achievable with a longer integration time. As pointed out in CCM21, with a noise similar to these experiments, constraints on the cosmological parameters remain similar to that of the Planck limit. Also, the frequency coverage of our hypothetical signal, especially on the higher frequency end, is wider than that of the EDGES or SARAS-3 experiment. The wider frequency coverage of our hypothetical signal is essential so that the signal be present during the EoR epoch (i.e., the 21~cm signal becomes sensitive to the reionization history), which will provide a more degeneracy-breaking potential to this signal while constraining different parameters. The input parameters (common to both reionization and 21~cm signal) used while generating the mock 21~cm signal are the best-fit values of different cosmological and reionization model parameters from the \textbf{CMB+Quasar} analysis. As is evident, the input values of other free parameters related to 21~cm estimation, which did not appear in the \textbf{CMB+Quasar} analysis, are $f_{Xh, *}$ and $f_{\alpha, *}$. To be consistent with the values estimated from the low redshift observations \citep{2006PhR...433..181F}, we take both $f_{X}$ and $f_{\alpha}$ to be equal to 1.0 and $f_{h}=0.2$. We take $f_*$ to be  $0.01$ consistent with our earlier works \citep[see,e.g.,][]{2015MNRAS.454L..76M,2018MNRAS.479.4566M,2021MNRAS.507.2405C}.  Note that with $f_{X}=1.0$, $f_{h}=0.2$, and $f_* =0.01$, the input parameter $f_{Xh, *}$ becomes 0.002 and $f_{\alpha, *}$ becomes 0.001. Since the actual value of $m_{X}$ is not known, we explore two scenarios to simulate the future data and for making the forecasts: one where the dark matter is the usual CDM (referred to as \textbf{CMB+Quasar+21cm$_{\mathbf{CDM}}$} hereafter) and another where $m_{X}= 7$ keV (referred to as \textbf{CMB+Quasar+21cm$_{\mathbf{7 keV WDM}}$} hereafter).

The ten free parameters for the joint analysis, including the 21~cm signal, are
\begin{equation*}
    \Theta =\{ H_{0}, \Omega_{b}h^{2}, \Omega_ch^{2}, A_{s}, n_s, f^{\mathrm{II}}_{\mathrm{esc}}, \lambda_{0}, m^{-1}_{X}, f_{Xh, *}, f_{\alpha, *} \}
\end{equation*}
In the presence of the mock 21~cm observations, the log-likelihood becomes
\begin{equation}\label{21cmlikelihood}
     \mathcal{L}=\mathcal{L}_{\mathrm{Pl}} + \mathcal{L}_{\mathrm{Re}} + \mathcal{L}_{21},
\end{equation}
Where $\mathcal{L}_{21}$ is the loglikelihood corresponding to mock observational data. Of course, the likelihood corresponding to 21~cm signal, $\mathcal{L}_{21}$, will depend on whether the mock data is generated with $m_{X} \xrightarrow[]{} \infty$ or $m_{X}=7$ keV as discussed below,

\subsubsection{\textbf{CMB+Quasar+21cm$_{\mathbf{CDM}}$}}

In this case, the likelihood term $\mathcal{L}_{21}$ in eqn-\ref{21cmlikelihood} will become
\begin{equation}
    \mathcal{L}_{21}= \sum_i \left[\frac{\delta T_b^{\mathrm{mock, CDM}}(\nu_i) - \delta T_b^{\mathrm{th}}(\nu_i)}{\sigma_i}\right]^2,
\end{equation}
where $\delta T_b^{\mathrm{mock, CDM}}(\nu_i)$ is the mock brightness temperature data generated using the CDM model.

\begin{figure*}
    \centering
    \includegraphics[width=\textwidth]{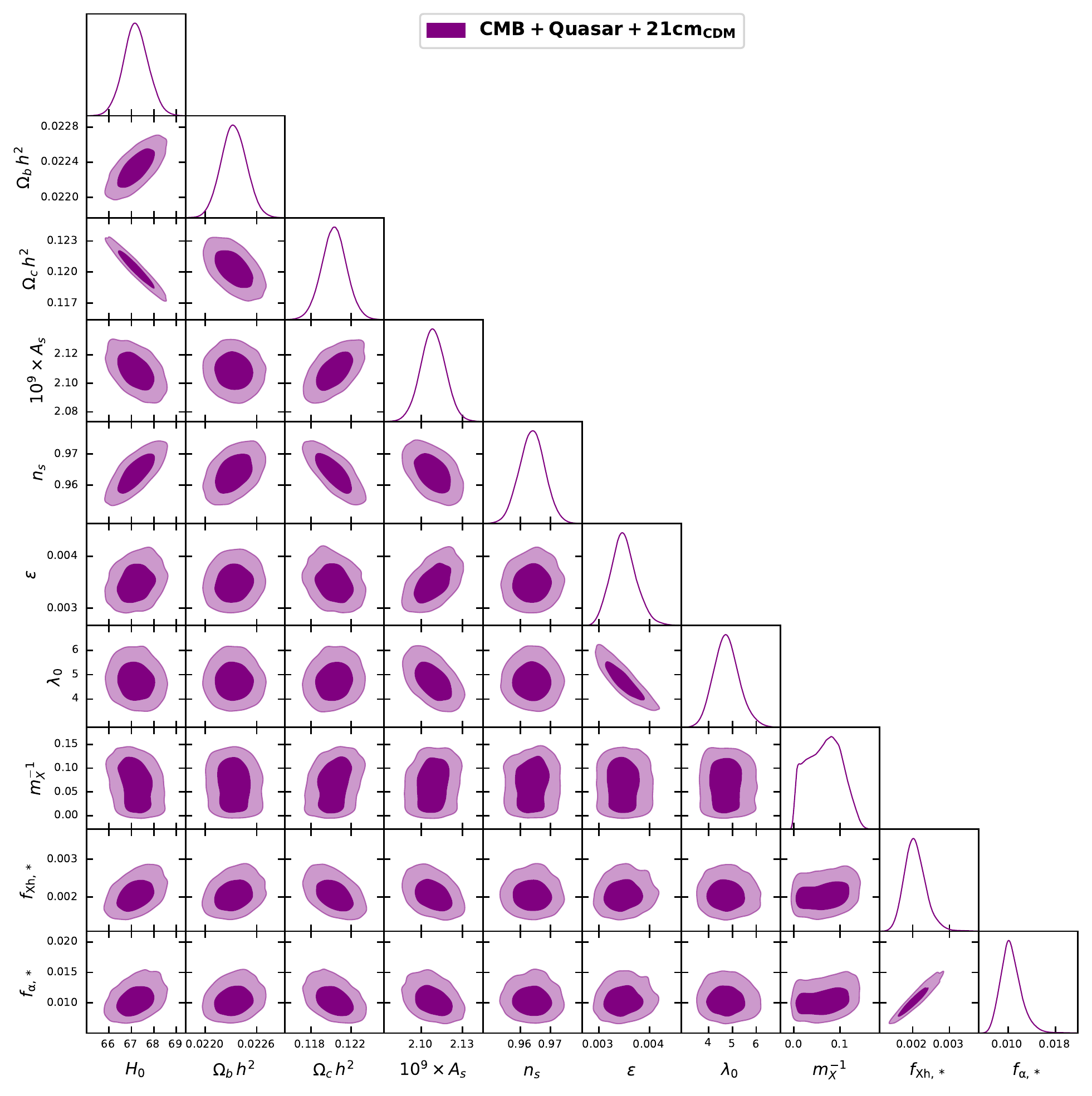}
    \caption{The marginalized posterior distribution of 10 free parameters obtained for the \textbf{CMB+Quasar+21cm$_{\mathbf{CDM}}$} case. Two-dimensional plots in the figure show the joint probability distribution (confidence contours at $68\%$ and $95\%$ ) of any two parameters.}
    \label{fig:CDM}
\end{figure*}

Once the MCMC run fulfils the convergence criteria, the 1D marginalized distribution of the quantity $\left(\frac{m_{X}}{keV}\right)^{-1}$ is shown (in magenta) in the figure-\ref{fig:compare_mx}. It is evident from this figure that the inclusion of 21 cm data forecasts the constraints to be $m_{X} >7.7$ keV ($95\%$ confidence level), which is much more stringent than that derived from the \textbf{CMB+Quasar} case. It is slightly weaker than the constraints coming from \cite{2021ApJ...917....7N} (they constrain $m_{X} <9.7$ keV) and is even stronger than the result obtained in \cite{PhysRevD.98.083540} (as mentioned earlier, their most stringent constraint on the WDM particles comes out to be $m_{X}< 4.1$ keV). Meanwhile, figure-\ref{fig:CDM} shows the constraints and the posterior distribution of all the free parameters used in this analysis. Unlike \textbf{CMB+Quasar} case, here, $m^{-1}_{X}$ does not show any correlation with any of the free parameters. This is because, in this case, the allowed range of values of $m_{X}$ is very high (i.e, $m_{X} >7.7$ keV ), and these high $m_{X}$ WDM models are practically indistinguishable from CDM. Note that in \textbf{CMB+Quasar} case, only the low value of $m_{X}$ shows the correlation/anti-correlation.

We also note from the 1-D posterior distribution of $f_{Xh, *}$ and $f_{\alpha, *}$ (bottom row of figure-\ref{fig:CDM}) that their best-fit values are 0.002 and 0.01 respectively. This is expected as those were the input values of these two parameters at the time of creating the hypothetical signal. This result also shows that our MCMC analysis with \texttt{CosmoReionMC} is excellent at recovering the ``true" parameters of the mock signal.

\subsubsection{\textbf{CMB+Quasar+21cm$_{\mathbf{7 keV WDM}}$}} For mock data generated with $m_{X}=7$ keV, the likelihood term $\mathcal{L}_{21}$ in eqn-\ref{21cmlikelihood} will become
\begin{equation}
    \mathcal{L}_{21}= \sum_i \left[\frac{\delta T_b^{\mathrm{mock, 7keV}}(\nu_i) - \delta T_b^{\mathrm{th}}(\nu_i)}{\sigma_i}\right]^2,
\end{equation}
where $\delta T_b^{\mathrm{mock, 7keV}}(\nu_i)$ is the mock brightness temperature data generated using 7keV WDM model.

After the completion of the MCMC run, the 1D posterior distribution of $(m_X/\rm{keV})^{-1}$ is shown in green in figure-\ref{fig:compare_mx}. As shown from the posterior distribution, the $95\%$ confidence level of $(m_X/\rm{keV})^{-1}$ comes out to be $[0.1, 0.2]$ implying that the future 21 cm data should allow detection of the WDM particles if $m_{X} \sim 7$ keV. 

\begin{figure*}
    \centering
    \includegraphics[width=\textwidth]{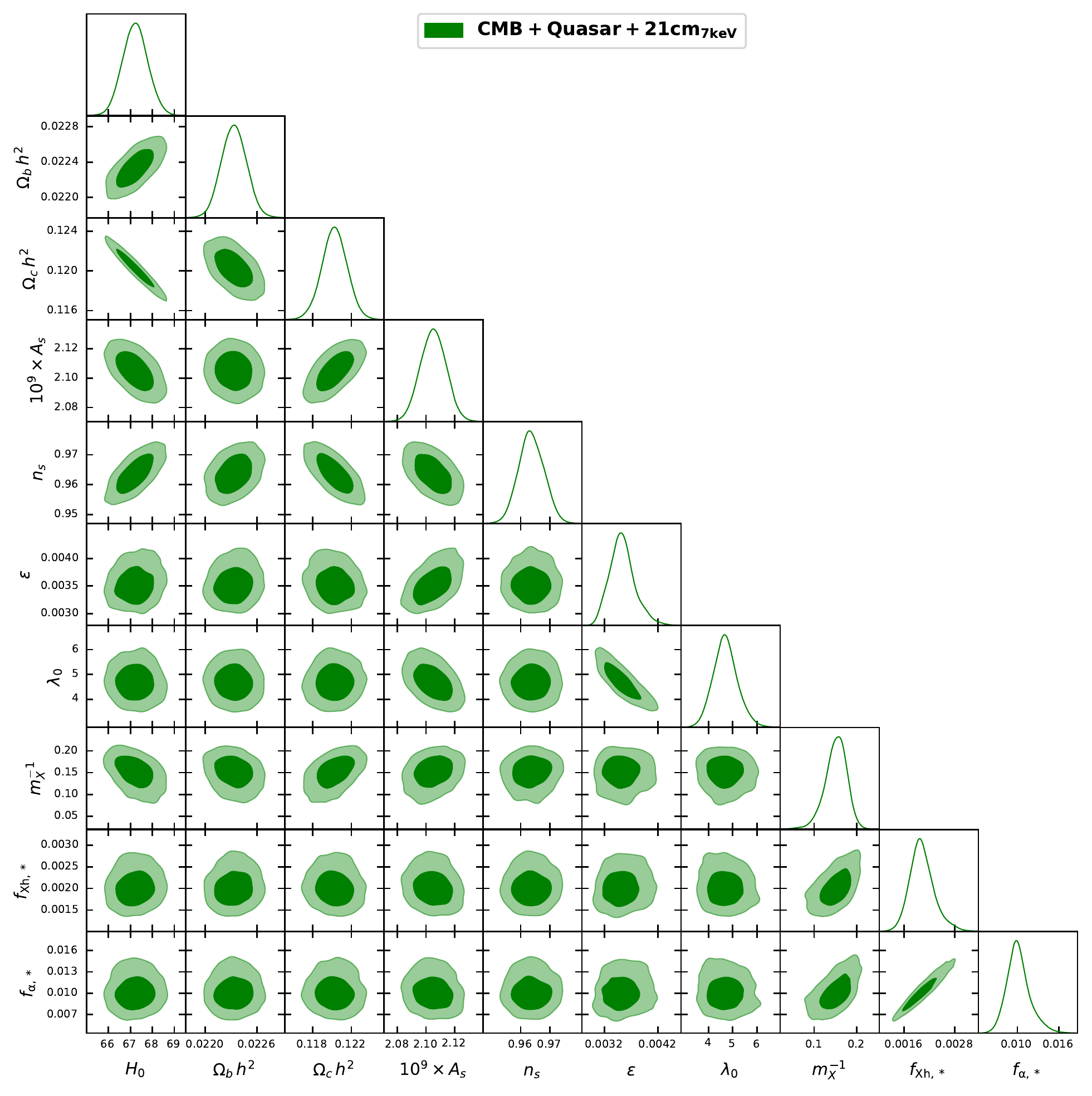}
    \caption{The marginalized posterior distribution of 10 free parameters obtained for the \textbf{CMB+Quasar+21cm$_{\mathbf{7 keV WDM}}$} case. Two-dimensional plots in the figure show the joint probability distribution (confidence contours at $68\%$ and $95\%$ ) of any two parameters.}
    \label{fig:7_kev}
\end{figure*}

The free parameters' posterior distribution is shown in figure-\ref{fig:7_kev}. It is evident that  $m^{-1}_{X}$ is strongly correlated with both $f_{Xh, *}$ and $f_{\alpha, *}$. It is because with smaller and smaller value of $m_{X}$ (higher and higher in $m^{-1}_{X}$), structure formation gets delayed making the appearance of absorption trough at later and later redshifts, the only way to keep both the redshift and the depth of the absorption trough unchanged is to increase the value of  $f_{Xh, *}$ and $f_{\alpha, *}$. It is also clear from the plot that $f_{Xh, *}$ and $f_{\alpha, *}$ are strongly correlated with each other. It is due to the fact that the increase in $f_{\alpha, *}$ will try to make the absorption trough deeper, and the only way to compensate for that is to increase the value of $f_{Xh, *}$.

\begin{figure*}
	\begin{subfigure}{0.4\textwidth} 
    \includegraphics[width=1.0\textwidth]{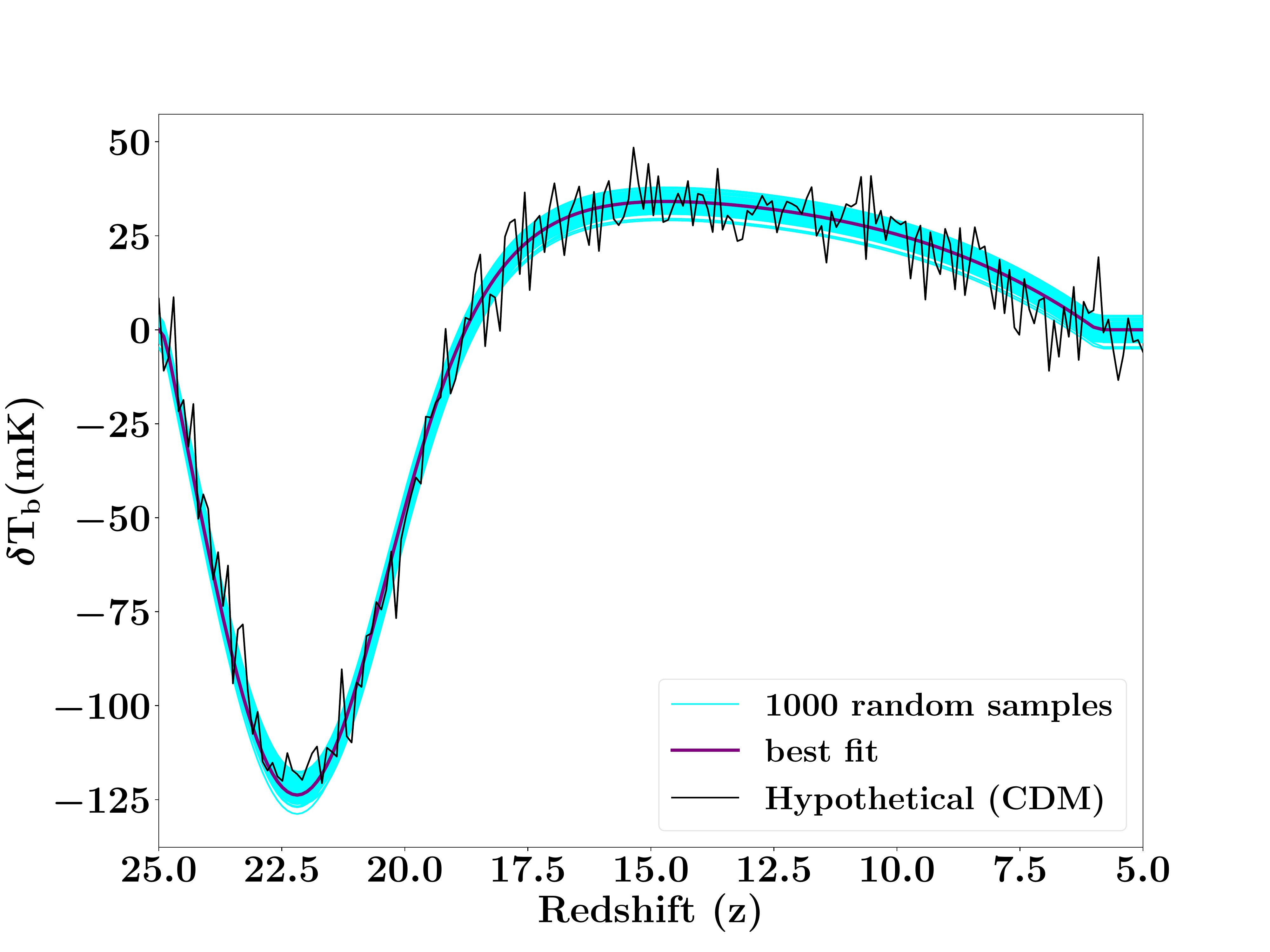}
	\end{subfigure}
	\begin{subfigure}{0.4\textwidth} 
    \includegraphics[width=1.0\textwidth]{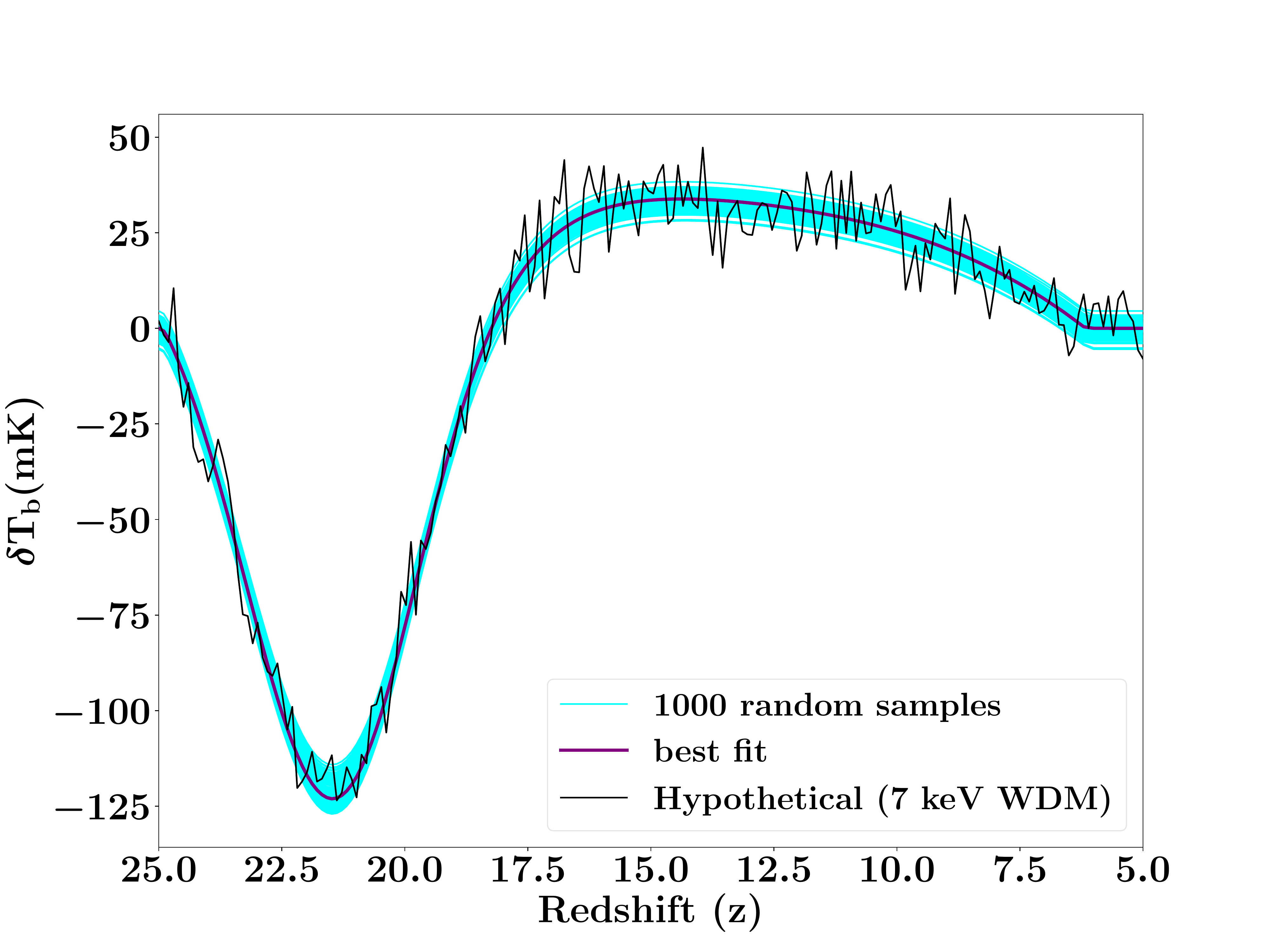}
	\end{subfigure}
	\hspace{1em} 
	\caption{Comparison between the mock 21~cm data and the signal recovered from the MCMC run. The left panel shows the case wh
 en the mock signal is produced from the CDM model, and the right panel represents the scenario with the mock data produced from the 7 keV WDM model. In both the panels, the black, magenta and cyan curves represent the mock data, the best-fit model and models corresponding to 1000 random samples from the MCMC chain, respectively.} 
\label{fig:compare_21cm} 
\end{figure*}

In figure-\ref{fig:compare_21cm}, we demonstrate the comparison between the mock signal and the recovered signal coming from the MCMC chain after its convergence. It is clear that the best-fit signal recovered from the MCMC run is in excellent agreement with the mock data in both cases.

\section{Conclusion and Discussion}\label{conclude}
In this work, with the help of our previously developed MCMC-based parameter estimation pipeline \texttt{CosmoReionMC}, we explored three different scenarios to constrain the mass of WDM particles and also shed light on the long-overlooked issue of the degeneracy between cosmological and astrophysical parameters while constraining the mass of the WDM particles. First, we demonstrate that when CMB and Quasar absorption-related observations are used along with the CMB angular power spectrum observations, the WDM particles with  $m_{X}<2.8$ keV can readily be ruled out. In the next step, we add a hypothetical 21~cm global signal along with the CMB and Quasar absorption-related observations to check if adding a 21~cm signal can put more stringent constraints on the mass of WDM compared to the already existing constraints. To this end, we generate two mock 21~cm signals, one with $m_{X} =7$ keV and the other assuming the usual CDM model (i.e., $m_{X} = \infty$). For the first scenario, the forecasts give $\left(m_X / \text{keV}\right)^{-1}$ in the range $[0.1, 0.2]$ ($95\%$ confidence level) implying that the future 21~cm data should allow detection of the WDM particles if $m_X \sim 7$~keV. In the second case, the inclusion of 21~cm data forecasts the constraints to be $m_X > 7.7$~keV ($95\%$ confidence level), much stronger than the present ones. 

Finally, we will discuss some of the caveats of our analysis presented here. First, we take all the free astrophysical efficiency parameters, e.g., the escape fraction, the X-ray heating efficiency, and the Ly$\alpha$ flux efficiency, to be constants i.e., they do not change with redshift or halo mass. But it is entirely possible that they are not constants as we have assumed here, and therefore the constraints on $m_X$ could be different from the analysis presented here. However, in spite of the simple assumptions, our work highlights the importance of the global 21~cm experiments in constraining the WDM particle mass.

In future work, we are planning to include more observational data e.g., UVLF data from the JWST observations \citep{2022ApJ...929....1H, 2023ApJS..265....5H, 2021AJ....162...47B, 2022arXiv221102607B, 2022ApJ...940L..14N} and then revisit the constraints on different parameters. We are also considering using a more accurate reionization model to eliminate some of the simplified assumptions used in the code. For example, we ignore molecular cooling completely in our model, despite the fact that molecular cooling for dark matter halos is an important mechanism that can change the PopIII star formation rate inside a halo.  In addition to that, we are also planning to include redshift/halo mass dependency in the efficiency parameters used in this analysis.

\section*{Acknowledgements}
 AC and TRC acknowledge the support of the Department of Atomic Energy, Government of India, under project no. 12-R\&D-TFR-5.02-0700. AC would also like to thank Aditya Chowdhury for his constant inspiration to write this paper.

\section*{Data Availability}
The observational data used here are taken from the literature and the code underlying this article will be shared on reasonable request to the corresponding author.


\bibliographystyle{mnras}
\bibliography{reion_WDM}

\appendix

\section{Dependence of mean free path of photons on the WDM particle mass}
\label{appendix: mfp}

\begin{figure*}
    \centering
    \includegraphics[width=\columnwidth]{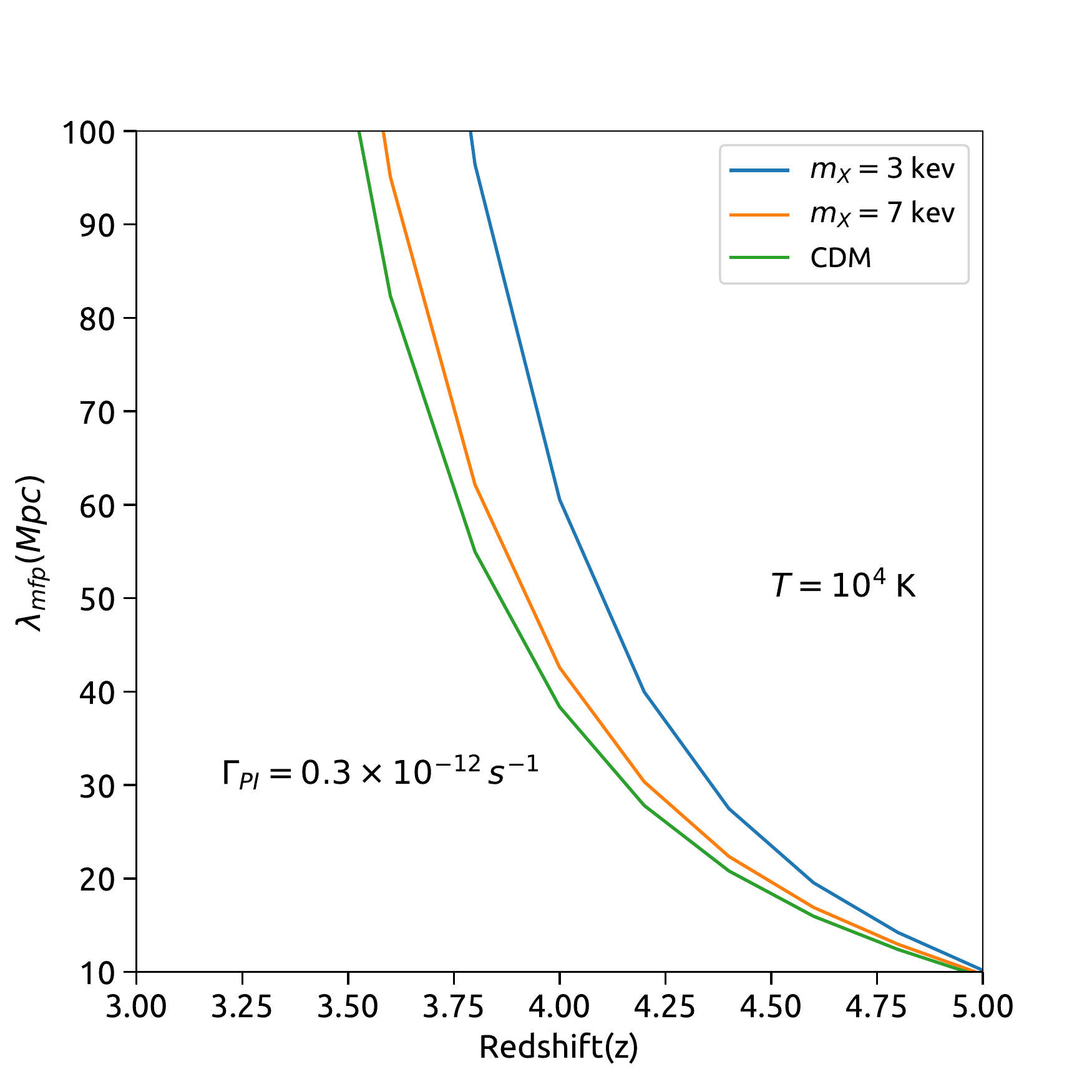}
    \caption{Redshift evolution of the mean free path of the ionizing photon ($\lambda_{\rm mfp}$) for CDM and WDM models with different particle mass. The green, orange and blue curves, respectively denotes CDM, 7keV and 3keV WDM. To make sure that the $\lambda_{\rm mfp}$ depends only on the density of the IGM, we kept the temperature of the IGM fixed at $10^4$ K and $\Gamma_{\rm PI} = 0.3 \times 10^{-12} {\rm sec^{-1}}$ in a completely ionized Universe. As is obvious, the mean free path $\lambda_{\rm mfp}$ decreases with increasing $m_{X}$.} 
    \label{fig:mfp}
\end{figure*}

We compare in Figure-\ref{fig:mfp} the dependence of the mean free path $(\lambda_{\rm mfp})$ on the mass of the WDM particle. For this exercise, we fix the hydrogen photoionization rate $\Gamma_{\rm PI}$ to a value of $0.3 \times 10^{-12} \, {\rm sec^{-1}}$ and the IGM temperature to $10^4 $ K for all the DM models. We study the redshift evolution of $\lambda_{\rm mfp}$ only in the redshift range where the Universe is fully ionized in all the DM models. Fixing these quantities ensures that any variation in $(\lambda_{\rm mfp})$ arising from differences in the ionization and thermal histories for the different DM models is absent, the mean free path depends only on the density distribution of the IGM. It is evident from the figure that the $\lambda_{\rm mfp}$ obtained in our analysis increases with the decreasing mass of the WDM particle. Recently, \cite{2023JCAP...01..002C} have presented similar analysis using Hydrodynamic simulations. We would like to point out that our result regarding the dependence of $\lambda_{\rm mfp}$ on the WDM particle mass, i.e., increase of $\lambda_{\rm mfp}$ with decrease in WDM particle mass, from our extremely efficient semi-analytic model indeed matches the overall trend obtained from their detailed hydrodynamical simulation.

\bsp	
\label{lastpage}
\end{document}